\documentclass[12pt]{iopart}
\usepackage{graphicx}

\begin{document}

\title[]{Relativistic quantum mechanical spin-1 wave equation in 2+1
dimensional spacetime}
\author{Mustafa Dernek, Semra Gurtas Dogan, Yusuf Sucu, Nuri Unal}

\begin{abstract}
In this study, we introduce a relativistic quantum mechanical wave equation
of the spin-1 particle as an excited state of the zitterbewegung and show
that it is consistent with the 2+1 dimensional Proca theory. At the same
time, we see that in the rest frame this equation has two eigenstates,
particle and antiparticle states or negative and positive energy
eigenstates, respectively, and satisfy $SO(2,1)$ spin
algebra. As practical applications, we derive the exact solutions of the
equation in the presence of a constant magnetic field and a curved
spacetime. From these solutions, we find Noether charge by integrating the
constructed spin-1 particle current on hyper surface and discuss pair
production from the charge. And, we see that the discussion on \ the Noether
charge is useful tool for undersdantding the pair production phenomenon
because the charge is derived from a probabilistic particle current.
\end{abstract}

\address{Department of Physics, Faculty of Science, Akdeniz University,
\\ 07058 Antalya, Turkey.} \ead{ysucu@akdeniz.edu.tr}
\vspace{10pt} 

\maketitle
\section{Introduction}

\label{sec:intro}

\qquad The physics in $2+1$ dimensional spacetime presents many interesting
and surprising results, both experimentally and theoretically. Therefore, in
recent years, $2+1$ dimensional theories have been widely studied in
physical areas, such as gravity, high energy particle theory, condensed
matter physics (e.g. monolayer structures), topological field theory, and
string theory \cite{a1,a2,a3,a4,a5,a6,a7,a8,a9,a10,a11}. In the view of
general relativity, 2+1 dimensional gravity has a number of solutions such
as black hole \cite{a12}, wormhole \cite{a13}, and propagating gravitational
wave \cite{a14,a15} .\emph{\ }On the other hand, in the view of the 2+1
quantum electrodynamics (QED$_{2+1}$), the properties of an electron in
graphene can be described by the 2+1 dimensional Dirac equation \cite%
{a16,a17,a18,a19}. Also, the inter-quark potential in quantum chromodynamics
is\emph{\ }studied in these dimensions and it has a rich structure, as in
the 3+1 dimensional spacetime \cite{a20,a21}. Additionally, in the QED$_{2+1}
$ context, massive or massless Dirac equation has provided important
physical results in flat \cite{b1,a23,a24,s1,s2} and curved backgrounds \cite%
{a25,a26,b3,b43,b4}. Although the physical properties of the Dirac particle
has been widely investigated in the context, the physical behavior of the
massive or massless spin-1 particle has not almost been considered, quantum
mechanically, in this framework, so far.

Duffin-Kemmer-Petiau equation as a relativistic quantum mechanical wave
equation for the spin-1 and spin-0\ particles has a long history \cite%
{b27,b28,b29,b30,b31,b32,b33}. The equation has been discussed in the 3+1
spacetime dimensions whether there exists or not a classical correspondence
in the context of Maxwell theory \cite{b34}. Also, the relativistic quantum
mechanical wave equation for the spin-1 particles is derived as an excited
state of zitterbewegung in the same dimensions, but it does not include
spin-0 part \cite{b35,b36,b37}. The massless case of the equation was
derived as a simple model of the zitterbewegung \cite{b38} and its
equivalence with the Maxwell equations in a flat and curved background were
discussed \cite{b38,b39}. The spin-1 equation was also solved in the
exponentially expanding universe and showed that its solutions are identical
with complexified Maxwell equations in the limit $m^{2}\rightarrow 0\ $ \cite%
{b40}.

The Maxwell theory is not valid in 2+1 dimensional spacetime because the
magnetic field becomes pseudoscalar. Therefore, in 2+1 dimensional
spacetime, the electrodynamical events are, classically, described by the
2+1 dimensional massive Proca gauge theory, which is called
Maxwell-Chern-Simons theory, and\ it is completely different from the
Maxwell theory. Also, contrary to even dimensional cases, in the
topologically massive gauge theories, the gauge fields, $F^{\nu \rho }$, and
potentials, $A_{\mu }$, are related to each others as follows:
\[
A_{\mu }=\frac{1}{2m}\varepsilon _{\mu \nu \rho }~F^{\nu \rho },
\]%
\cite{c41,c42,c43,c44,c45,c46} and, at the same time, the spin-1 particle
wave equation was discussed in this context. Also, with the pseudoclassical
approach, the spin-1 particle wave equation in the 2+1 dimensions was
studied by canonical quantization \cite{c47,c48,c49}. Nevertheless, it has
still been continued some discussions on such an equation in the 2+1
dimensional spacetime structure, physically and mathematically \cite%
{d1,d2,d3,m6}. On all these points of view, a relativistic quantum
mechanical wave equation for a spin-1 particle in the 2+1 dimensional
spacetime should be expected to satisfy the Proca equation and the relation
between vector potentials,$\ A_{\mu }$, and tensor fields,$\ F^{\mu \nu }$,
in classical limit. With these motivations, we discuss, in detail, a
relativistic quantum mechanical wave equation for a spin-1 particle directly
derived from the 2+1 dimensional Barut-Zanghi model \cite{b35} and find a
relation between this equation and the 2+1 dimensional Proca theory. As a
practical application, we also discuss the exact solutions of the wave
equation in a constant magnetic field and a curved background. Moreover, we
write a current expression of the spin-1 particle in this context and
calculate it from these solutions. And, finally, we find the Noether charge
derived from the current and discuss pair production in terms of the charge.
And, we see that such a discussion is useful tool for understanding the
pair production phenomenon since this charge is derived from a probabilistic
particle current.

The outline of this study is as follows: In Section 2, in the 2+1
dimensional spacetime, we introduce a relativistic quantum mechanical wave
equation for a spin-1 particle as an excited state of the classical
zitterbewegung model and construct its free particle solutions. We also
discuss the particle and antiparticle solutions of the equation in the rest
frame and show that the equation is equivalent to the Proca equation by
defining vector potentials and electromagnetic fields in terms of the
components of the spin-1 particle spinor. In Section 3, we obtain the exact
solutions of the spin-1 particle equation in a constant magnetic field. From
these solutions we derive the energy eigenvalues of the particle and write
the current densities and Noether charge. In Section 4, we find the exact
solutions of the equation in the contracting and expanding 2+1 dimensional
curved background. And, we derive the current and Noether charge from these
solutions. In Conclusion, we evaluate the results of the study.

\section{The spin-1 particle in the 2+1 dimensional flat spacetime}

\label{sec: b1}

\qquad A relativistic quantum mechanical wave equation for the spin-1
particle introduced in the 3+1 dimensions was discussed as an excited state
of the classical zitterbewegung model \cite{b35,b36,b37}. As the classical
model of the zitterbewegung \cite{b35} in the 2+1 dimensional spacetime, for
which its symmetry and integrability properties are investigated \cite{c50},
is quantized, it directly gives us the following relativistic quantum
mechanical wave equation for the evolution of the free spin -1 particle \cite%
{b36}:
\begin{equation}
\left\{ \left( \overline{\mathbf{\sigma }}^{\mu }\otimes 1+1\otimes
\overline{\mathbf{\sigma }}^{\mu }\right) P_{\mu }-\left( 1\otimes 1\right)
2M\right\} _{\alpha ^{\prime }\alpha }\Psi _{\alpha \beta }\left( x\right)
=0,  \label{1}
\end{equation}%
where $M$ and $P_{\mu }$ are the mass and the energy-momentum of the
particle, respectively, and $\overline{\mathbf{\sigma }}^{\mu }$ is defined
in terms of the Pauli matrices as $\overline{\mathbf{\sigma }}^{\mu }=\left(
\sigma ^{3},~i\sigma ^{1},~i\sigma ^{2}\right) ,$ and $1$ is unit matrix.
The matrices $\overline{\mathbf{\sigma }}^{\mu }$ satisfy anti-commutation
relations in 2+1 dimensional spacetime:%
\[
\left\{ \overline{\mathbf{\sigma }}_{\mu },\overline{\mathbf{\sigma }}%
_{\upsilon }\right\} =2\eta _{\mu \nu },
\]%
where $\eta _{\mu \nu }=$diag$(1,-1,-1)$, is Minkowski metric in 2+1
dimensional spacetime. The spinor of the spin-1 particle, $\Psi
_{\alpha\beta }\left( x\right) ,$ is defined as
\begin{equation}
\Psi _{\alpha \beta }\left( r,t\right) =\left( \psi _{+}~,~\psi _{0}~,~\psi
_{0~},~\psi _{-}\right) ^{T}.  \label{2}
\end{equation}
where the $T$ means transpose of the row matrix. In this model, the wave
function, $\Psi _{\alpha \beta }\left( x\right) ,$ is the symmetric spinor
with rank 2 and it is represented as a direct product of two Dirac spinors
in the 2+1 dimensions and the quantization of the classical system requires
that $\Psi _{\alpha \beta }\left( x\right) $ should be symmetric with
respect to the indices $\alpha $,\ $\beta ,$ where the first (second)
indices correspond to the first (second) set of the Dirac matrices. For this
reason, $\Psi _{\alpha \beta }\left( x\right) $ has three independent
components, $\psi _{+}$, $\psi _{0}$, $\psi _{-}$.

If we choose a plane wave solutions as follows;
\begin{equation}
\Psi \left( \mathbf{r},t\right) =\left(
\begin{array}{c}
\phi _{+} \\
\phi _{0} \\
\phi _{0} \\
\phi _{-}%
\end{array}%
\right) e^{-ip_{\mu }x^{\mu }},  \label{3}
\end{equation}%
then, the explicit form of the spin-1 particle equation is written in terms
of the spinor components, $\phi _{+},$ $\phi _{0},$ $\phi _{-}:$%
\begin{eqnarray}
\left( P_{0}-M\right) \phi _{+}~+\left( iP_{1}+P_{2}\right) \phi _{0}=0,
\nonumber \\
\left( iP_{1}-P_{2}\right) \phi _{+}+\left( iP_{1}+P_{2}\right) \phi
_{-}-2M\phi _{0} =0,  \label{4} \\
\left( P_{0}+M\right) \phi _{-}~-\left( iP_{1}-P_{2}\right) \phi _{0} =0.
\nonumber
\end{eqnarray}%
If we add and subtract the first and third rows and organize the second row
of Eq (\ref{4}), the following equations are obtained, respectively,
\begin{eqnarray}
P_{0}\left( \phi _{+}-\phi _{-}\right) +iP_{1}\left( 2\phi _{0}\right)
=M\left( \phi _{+}+\phi _{-}\right) ,  \nonumber \\
P_{0}\left( \phi _{+}+\phi _{-}\right) +P_{2}\left( 2\phi _{0}\right)
=M\left( \phi _{+}-\phi _{-}\right) ,  \label{5} \\
iP_{1}\left( \phi _{+}+\phi _{-}\right) -P_{2}\left( \phi _{+}-\phi
_{-}\right) =M\left( 2\phi _{0}\right) .  \nonumber
\end{eqnarray}%
And, we reorganize Eq (\ref{5}) as a first order partial differential equations
system:
\begin{eqnarray}
\partial ^{0}\left( i\frac{\phi _{+}-\phi _{-}}{\sqrt{M}}\right) -\partial
^{1}\left( \frac{2\phi _{0}}{\sqrt{M}}\right) =\sqrt{M}\left( \phi _{+}+\phi
_{-}\right) ,  \nonumber \\
\partial ^{0}\left( -\frac{\phi _{+}+\phi _{-}}{\sqrt{M}}\right) -\partial
^{2}\left( \frac{2\phi _{0}}{\sqrt{M}}\right) =\ i\sqrt{M}\left(
\phi_{+}-\phi _{-}\right) ,  \nonumber \\
\partial ^{1}\left( -\frac{\phi _{+}+\phi _{-}}{\sqrt{M}}\right) -\partial
^{2}\left( i\frac{\phi _{+}-\phi _{-}}{\sqrt{M}}\right) =\sqrt{M}\left(
2\phi _{0}\right) .  \label{6}
\end{eqnarray}%
From this equation system, if we define the complex gauge potentials and
fields in terms of the spinor components as
\begin{equation}
A^{0}=\frac{2\phi _{0}}{\sqrt{M}},\ \ A^{1}=i\frac{\phi _{+}-\phi _{-}}{%
\sqrt{M}},\ \ A^{2}=-\frac{\phi _{+}+\phi _{-}}{\sqrt{M}}  \label{7}
\end{equation}%
and
\begin{equation}
F^{01}=\sqrt{M}\left( \phi _{+}+\phi _{-}\right) ,\ \ F^{02}=i\sqrt{M}\left(
\phi _{+}-\phi _{-}\right) ,\ \ F^{12}=\sqrt{M}\left( 2\phi _{0}\right) ,
\label{8}
\end{equation}%
respectively, then, using the relations given Eqs (\ref{7}) and (\ref{8}),
we see that Eq (\ref{6}) satisfies the well known relations between the
gauge fields and potentials as follows;%
\begin{equation}
\partial ^{\mu }A^{\nu }-\partial ^{\nu }A^{\mu }=F^{\mu \nu }.  \label{9}
\end{equation}%
According to the definitions in Eqs (\ref{7}) and (\ref{8}), then, Eq (\ref{6}%
) becomes;
\begin{eqnarray}
i\sqrt{M}\partial _{0}\left( \phi _{+}-\phi _{-}\right) +\sqrt{M}\partial
_{1}\left( 2\phi _{0}\right) &=&M^{2}\left( \frac{\phi _{+}+\phi _{-}}{\sqrt{%
M}}\right) ,  \nonumber \\
i\sqrt{M}\partial _{0}\left( \phi _{+}+\phi _{-}\right) -i\sqrt{M}\partial
_{2}\left( 2\phi _{0}\right) &=&M^{2}\left( \frac{\phi _{+}-\phi _{-}}{\sqrt{%
M}}\right) ,  \label{10} \\
\sqrt{M}\partial _{1}\left( \phi _{+}+\phi _{-}\right) +i\sqrt{M}\partial
_{2}\left( \phi _{+}-\phi _{-}\right) &=&M^{2}\left( \frac{2\phi _{0}}{\sqrt{%
M}}\right),  \nonumber
\end{eqnarray}%
and these equations are in the form of the massive Proca equation in the 2+1
dimensional spacetimes:%
\begin{equation}
\partial _{\mu }F^{\mu \nu }+M^{2}A^{\nu }=0.  \label{11}
\end{equation}%
In particular, from the definitions in Eqs (\ref{7}) and (\ref{8}), we
notice that the expressions of the vector potentials and fields in Eqs(\ref%
{7}) and (\ref{8}), respectively, are related to each other by the following
relations;
\begin{equation}
A^{\mu }=\frac{1}{2M}\epsilon ^{\mu \nu \rho }F_{\nu \rho },  \label{12}
\end{equation}%
where $\epsilon ^{\mu \nu \rho }$ is the Levi-Civita symbol. These results
are consistent with Ref. \cite{c46}, and also if we eliminate the potentials
by using Eq (\ref{12}) in the Proca equation Eq (\ref{11}), then, we derive
the following equation;
\begin{equation}
\partial _{\mu }F^{\mu \nu }+\frac{M}{2}\epsilon ^{\nu \alpha \beta
}F_{\alpha \beta }=0.  \label{13}
\end{equation}%
Therefore, we say that these equations are compatible with the results of
the topologically massive gauge theories\emph{\ }\cite{c43,c44,c45,c46}.

To discuss the free particle solutions of the spin-1 particle from Eq (\ref%
{4}), at first, the components $\phi _{+}$ and $\phi _{-}$ are written as%
\begin{equation}
\phi _{\pm }=\frac{P}{P_{0}\mp M}e^{\mp i\left( \varphi +\frac{\pi }{2}%
\right) }\phi _{0},  \label{14}
\end{equation}%
where $\tan \varphi =P_{2}/P_{1}$ and $P$ is the magnitude of the momentum
vector, $\mathbf{P}=\left( P_{1},~P_{2}\right) $. Then, substituting $\phi
_{+}$ and $\phi _{-}$ into the second row of Eq (\ref{4}), we obtain the
following free particle relativistic wave equation for $\phi _{0}$;
\begin{equation}
\left( P_{0}^{2}-P_{1}^{2}-P_{2}^{2}-M^{2}\right) \phi _{0}=0.  \label{15}
\end{equation}%
This is the Proca equation for $\phi _{0}$ in the 2+1 dimensions and it has
the correct energy-momentum condition. Then, the normalized wave function
for Eq (\ref{3}) is obtained as%
\begin{equation}
\Psi \left( \mathbf{r},t\right) =\frac{1}{2\sqrt{\left\vert P_{0}\right\vert
M}}\left(
\begin{array}{c}
\left( M+P_{0}\right) e^{-i\left( \varphi +\frac{\pi }{2}\right) } \\
P \\
P \\
\left( M-P_{0}\right) e^{i\left( \varphi +\frac{\pi }{2}\right) }%
\end{array}%
\right) e^{-ip_{\mu }x^{\mu }},  \label{16}
\end{equation}%
where the normalization condition is given by%
\begin{equation}
\Psi \left( \mathbf{r},t\right) ^{\ast T}\Psi \left( \mathbf{r},t\right) =%
\frac{\left\vert P_{0}\right\vert }{M}.  \label{17}
\end{equation}%
In the rest frame, the particle has only two states which are
\begin{equation}
\Psi \left( t\right) =\left(
\begin{tabular}{l}
$1$ \\
$0$ \\
$0$ \\
$0$%
\end{tabular}%
\ \right) e^{-iMt},\quad {for}\quad P_{0}=M  \label{18}
\end{equation}%
and
\begin{equation}
\Psi \left( t\right) =\left(
\begin{tabular}{l}
$0$ \\
$0$ \\
$0$ \\
$1$%
\end{tabular}%
\ \right) e^{iMt},\quad {for}\quad P_{0}=-M.  \label{19}
\end{equation}%
They are the particle and antiparticle solutions and, at same time, they are
eigenstates of spin operator, $S_{12},$ with eigenvalues $+1$ and $-1$,
respectively.

On the other hand, Eq (\ref{4}) can also be rewritten\ as matrix equation in
the following form:%
\[
\left( \beta ^{\mu }P_{\mu }-M\right) \Psi =0,
\]%
where the complex spinor$\ \Psi $ has the three components in the
representation:%
\[
\Psi \left( x_{1,}x_{2,}t\right) =\left( \psi _{+}~,\sqrt{2}~\psi
_{0}~,~\psi _{-}\right) ^{T},
\]%
and $\beta $ matrices are Hermitian spin-1 matrices in the 2+1 dimensional
spacetime and they can be defined as
\begin{equation}
\beta ^{0}=\resizebox{.13\hsize}{!}{$\left( \begin{array}{ccc} 1 & 0 & 0 \\
0 & 0 & 0 \\ 0 & 0 & -1 \end{array}\right)$} ,~\beta ^{1}=\frac{1}{\sqrt{2}}i%
\resizebox{.13\hsize}{!}{$\left( \begin{array}{ccc} 0 & 1 & 0 \\ 1 & 0 & 1
\\ 0 & 1 & 0\end{array}\right)$} ,~\beta ^{2}=\frac{1}{\sqrt{2}}%
\resizebox{.13\hsize}{!}{$\left( \begin{array}{ccc} 0 & 1 & 0 \\ -1 & 0 & 1
\\ 0 & -1 & 0\end{array}\right)$}.  \label{20}
\end{equation}%
And, it is clear that they satisfy the following $SO(2,1)$\emph{\ }spin-1
algebra:%
\[
\left[ \beta ^{\mu },\beta ^{\nu }\right] =-i\epsilon ^{\mu \nu \rho }\beta
_{\rho }.
\]%
These matrices also satisfy the following relation:%
\[
\eta _{\mu \nu }\beta ^{\mu }\beta ^{\nu }=2I.
\]

To discuss the conserved current for the spin-1 particle, we point out that
the Hermitian conjugate of the $\beta ^{\mu }$ matrices obeys the following
transformation:%
\[
\left( \beta ^{\mu }\right) =\gamma \left( \beta ^{\mu }\right) ^{T\ast
}\gamma ,
\]%
where $\gamma $ is%
\begin{equation}
\gamma =\left(
\begin{array}{ccc}
1 & 0 & 0 \\
0 & -1 & 0 \\
0 & 0 & 1%
\end{array}%
\right) ,  \label{21}
\end{equation}%
and the Hermitian conjugate of the wave function is%
\begin{equation}
\overline{\Psi }=\left( \Psi \right) ^{T\ast }\gamma .  \label{22}
\end{equation}%
Then, in this context, the conserved current for the spin-1 particle becomes%
\begin{equation}
J^{\mu }=\overline{\Psi }\beta ^{\mu }\Psi ,  \label{23}
\end{equation}%
and the current components in terms of the spin-1 particle spinor components
are explicitly given as follows:
\begin{eqnarray}
J^{0} &=&\left\vert \phi _{+}\right\vert ^{2}-\left\vert \phi
_{-}\right\vert ^{2},  \nonumber \\
J^{1} &=&i\left( \phi _{+}+\phi _{-}\right) ^{\ast }\sqrt{2}\phi _{0}-i\sqrt{%
2}\phi _{0}^{\ast }\left( \phi _{+}+\phi _{-}\right) ,  \label{24} \\
J^{2} &=&\left( \phi _{+}-\phi _{-}\right) ^{\ast }\sqrt{2}\phi _{0}+\sqrt{2}%
\phi _{0}^{\ast }\left( \phi _{+}-\phi _{-}\right) .  \nonumber
\end{eqnarray}%
By means of the definitions in Eqs (\ref{7}) and (\ref{8}), the components
of the current may also be written in terms of the Proca fields as%
\begin{eqnarray}
J^{0} =\frac{1}{4M}\left[ \left\vert F^{01}-iF^{02}\right\vert
^{2}-\left\vert F^{01}+iF^{02}\right\vert ^{2}\right] ,  \label{25} \\
J^{1}+iJ^{2} =\frac{i}{4M}\left[ \left( F^{01}-iF^{02}\right) ^{\ast
}F^{12}-\left( F^{12}\right) ^{\ast }\left( F^{01}+iF^{02}\right) \right] ,
\label{26}
\end{eqnarray}%
where $J^{0}$ corresponds to the difference between the right handed and
left handed electric fields, $J^{1}$ and $J^{2}$ are the generalization of
the Poynting vectors into the 2+1 dimensional complex fields. Furthermore,
the current components can be expressed in terms of the gauge potentials and
fields as follows;%
\begin{eqnarray}
J^{0} &=&\frac{i}{4}\left[ \left( A_{1}^{\ast }F^{10}-A_{1}\left(
F^{10}\right) ^{\ast }\right) +\left( A_{2}^{\ast }F^{20}-A_{2}\left(
F^{20}\right) ^{\ast }\right) \right] ,  \label{27} \\
J^{1} &=&\frac{-i}{4}\left[ \left( A_{0}^{\ast }F^{01}-A_{0}\left(
F^{01}\right) ^{\ast }\right) +\left( A_{2}^{\ast }F^{21}-A_{2}\left(
F^{21}\right) ^{\ast }\right) \right] ,  \label{28} \\
J^{2} &=&\frac{-i}{4}\left[ \left( A_{0}^{\ast }F^{02}-A_{0}\left(
F^{02}\right) ^{\ast }\right) +\left( A_{1}^{\ast }F^{12}-A_{1}\left(
F^{12}\right) ^{\ast }\right) \right] .  \label{29}
\end{eqnarray}
In particular, it is also important that these results are consistent with
the following expression for the conserved current of the Proca Fields in
3+1 dimensions \cite{c51,c52,c53,c54}:
\[
J_{\mu }=\varphi ^{\nu }G_{\mu \nu }^{\ast }-G_{\mu }^{\nu }\varphi _{\nu
}^{\ast }.
\]

\section{The spin-1 particle in a constant magnetic field}

\label{ns}

\qquad The motion of charged particles in a magnetic field is an important
problem in classical and quantum electrodynamics. Moreover, in 3+1
dimensions the problem of a charged particle in a constant magnetic field is
a 2+1 dimensional problem because of the axial symmetry \cite{m1, m2}.
Therefore, to discuss the relativistic quantum mechanical behavior of the
charged spin-1 particle in a constant magnetic field, we write the
relativistic wave equation for the spin-1 particle in the presence of
electromagnetic fields as

\begin{equation}
\left[ \left( \sigma ^{\mu }\otimes I+I\otimes \sigma ^{\mu }\right) \pi
_{\mu }-2(I\otimes I)M\right] \Psi =0,  \label{30}
\end{equation}%
where $\pi _{\mu }$ is the generalized momentum of the particle and the
explicit form of its in terms of an electromagnetic potential, $A_{\mu },$
is $\pi _{\mu }=p_{\mu }-eA_{\mu }$, where $e$ is the charge of the
particle. Then, Eq (\ref{30}) becomes
\begin{equation}
\left[
\begin{array}{cccc}
2(\pi _{0}-M) & i(\pi _{1}-i\pi _{2}) & i(\pi _{1}-i\pi _{2}) & 0 \\
i(\pi _{1}+i\pi _{2}) & -2M & 0 & i(\pi _{1}-i\pi _{2}) \\
i(\pi _{1}+i\pi _{2}) & 0 & -2M & i(\pi _{1}-i\pi _{2}) \\
0 & i(\pi _{1}+i\pi _{2}) & i(\pi _{1}+i\pi _{2}) & -2(\pi _{0}+M)%
\end{array}%
\right] \left[
\begin{array}{c}
\psi _{+} \\
\psi _{0} \\
\psi _{0} \\
\psi _{-}%
\end{array}%
\right] =0.  \label{31}
\end{equation}%
Letting $\pi _{\pm }=\pi _{1}\pm i\pi _{2},$ and writing $\pi _{0},$ and $%
\pi _{\pm }$ in polar coordinates, $\left( r,\theta \right) ,$ for a
constant magnetic field,i.e, $A_{0 }=0,$ $A_{1 }=0,$ $A_{2 }$=Br/2 as
follows;
\[
\pi _{0}=iE,
\]
\[
\pi_{\pm }=M\zeta e^{-i\theta}(-i\frac{\partial }{\partial\sqrt{\rho }}\pm
\frac{1}{\sqrt{\rho }}\frac{\partial }{\partial \theta }\mp i\sqrt{\rho }),
\]%
where $B$\ is magnitude of the constant magnetic field $\zeta=\sqrt{\frac{eB%
}{2M^{2}}}$, $\epsilon=\frac{E}{M}$, $\rho =\frac{{eB}r^{2}}{{2}}$ . Using separation of variable method, the components of the general wave
function can be written as follows;%
\begin{eqnarray}
\psi _{+}(\rho,\theta ) &=&\exp (-iEt)\exp (i(k+1)\theta )\varphi _{1}(\rho
),  \nonumber \\
\psi _{-}(\rho,\theta ) &=&\exp (-iEt)\exp (i(k-1)\theta )\varphi _{-1}(\rho
),  \label{35} \\
\psi _{0}(\rho,\theta ) &=&\exp (-iEt)\exp (ik\theta )\varphi _{0}(\rho ).
\nonumber
\end{eqnarray}%
And substituting the expressions of $\pi _{0},$ and $\pi _{\pm }$ and the
components of the general wave function into Eq (\ref{31}), it becomes
\[
\frac{(\epsilon-1)}\zeta\varphi _{1}(\rho )+2\sqrt{\rho }\left( \frac{d}{%
d\rho }+\frac{k}{2\rho }-\frac{1}{2}\right) \varphi _{0}(\rho )=0,
\]
\[
\frac{(\epsilon+1)}\zeta\varphi _{-1}(\rho )-2\sqrt{\rho }\left( \frac{d}{%
d\rho }-\frac{k}{2\rho }+\frac{1}{2}\right) \varphi _{0}(\rho )=0,
\]%
\begin{eqnarray}
\resizebox{.78\hsize}{!}{$\zeta\sqrt{\rho }[( \frac{d}{d\rho
}-\frac{1}{2\rho }( k-1) +\frac{1}{2}) \varphi _{1}(\rho )+( \frac{d}{d\rho
}+\frac{1}{2\rho }(k+1)-\frac{1}{2}) \varphi _{-1}(\rho)]=\varphi _{0}(\rho
)$}.  \label{37}
\end{eqnarray}%
Letting $\varphi _{0}(\rho )=\frac{1}{\sqrt{\rho }}\chi (\rho )$, we get the
Whittaker differential equation:
\begin{equation}
\frac{d^{2}\chi (\rho )}{\rho ^{2}}+\left( -\frac{1}{4}+\frac{\lambda }{\rho
}+\frac{\frac{1}{4}-\frac{k^{2}}{4}}{\rho ^{2}}\right) \chi (\rho )=0.
\label{39}
\end{equation}%
Then, we can construct the general solution in terms of the Whittaker
functions, $M_{\lambda ,\frac{k}{2}}(\rho )$, as follows;
\begin{eqnarray}
\left(
\begin{array}{c}
\psi _{+} \\
\psi _{0} \\
\psi _{-}%
\end{array}%
\right) =\resizebox{.62\hsize}{!}{$Ne^{i\left( k\theta -Et\right) }\left(
\begin{array}{c} \frac{-2\zeta e^{i\theta }}{\epsilon-1}\left[ \left(
\frac{k}{\rho } -1\right) M_{\lambda ,\frac{k}{2}}(\rho )+\frac{\left(
\frac{k+1}{2}-\lambda \right) }{k+1}M_{\lambda ,\frac{k}{2}+1}(\rho )\right]
\\ M_{\lambda ,\frac{k}{2}}(\rho )/\sqrt{\rho } \\ \frac{2\zeta e^{-i\theta }}{(\epsilon+1)}\frac{\left( \frac{k+1}{2}-\lambda \right)
}{k+1}M_{\lambda ,\frac{k}{2}+1}(\rho )\end{array}\right)$} ,
\end{eqnarray}
where $\lambda-\frac{k}{2} =\frac{\epsilon^{2}-1}{4\zeta^{2}}-\frac{\epsilon%
}{2}=n+\frac{1}{2}$, and $N$ is the normalization constant. From this
solution, the energy eigenvalues are obtained as
\begin{equation}
E_{\pm}=M(\zeta^{2}\pm \sqrt{\zeta^{4}+1+2\zeta^{2}\left( 2n+1\right)}),
\label{42}
\end{equation}
Letting the asymptotic form of the solutions in Eq (\ref{35}) \cite{m11}, we
get the components of the current in Eq (\ref{23}) as follows;
\begin{equation}
\resizebox{.71\hsize}{!}{$J^{0}=\left\vert N\right\vert ^{2}\frac{\Gamma
(k+1)^{2}}{\Gamma (n+k+1)^{2}}{\frac{4}{M}\zeta^{2}\rho ^{2n+k-1}\left[
\left( \frac{n+1+\rho }{\epsilon-1}\right) ^{2}-\left(
\frac{n}{\epsilon+1}\right) ^{2}\right] }e^{-\rho }$},  \label{j0}
\end{equation}
\begin{equation}
\resizebox{.73\hsize}{!}{$J^{1}+iJ^{2}=\left\vert N\right\vert
^{2}\frac{\Gamma (k+1)^{2}}{\Gamma (n+k+1)^{2}}{4\zeta M \rho
^{2n+k-\frac{1}{2}}\left[ 2\epsilon n+(\epsilon+1)(1+\rho )\right]
}e^{-i\theta }e^{-\rho }$},  \label{j1}
\end{equation}
where

\begin{eqnarray}
\vert N \vert& =\frac{\Gamma (n+k+1)}{\Gamma (2n+k)^{1/2}\Gamma (k+1)}
\nonumber  \label{j1..} \\
&\resizebox{.69\hsize}{!}{$\times\frac{M(\epsilon^{2}-1)}{((%
\epsilon+1)^{2}[(k+1)^{2}-4+(2n+k+1)(18n+3k+7)]-(%
\epsilon-1)^{2}4(n+k+1)^{2})^{1/2}}$}.
\end{eqnarray}
Under the strong magnetic field, i.e $\zeta^{4}\gg 1$, the energy eigenvalues Eq (\ref{42})
and the zeroth component of the current Eq (\ref{j0}) become, respectively,

\begin{equation}
E_{+}\approx 2M\zeta ^{2}+M(2n+1)+O{(\zeta ^{-2})},  \label{42a}
\end{equation}%
\begin{equation}
E_{-}\approx -M(2n+1)+O{(\zeta ^{-2})},  \label{42aa}
\end{equation}%
and
\begin{eqnarray}
J_{+}^{0}&=\frac{2M{\ \zeta ^{2}}}{\pi \Gamma (2n+k)}  \nonumber \\
&\resizebox{.62\hsize}{!}{$\times\frac{e^{-\rho }\rho ^{2n+k-1}\left[
\left( n+1+\rho \right) ^{2}- n^{2}\right]
}{[(k+1)^{2}-4+(2n+k+1)(18n+3k+7)-4(n+k+1)^{2}]}$},
\end{eqnarray}%
\label{j01}
\begin{eqnarray}
J_{-}^{0}&=\frac{2M{\ \zeta ^{2}}}{\pi \Gamma (2n+k)}  \nonumber \\
&\resizebox{.66\hsize}{!}{$\times\frac{e^{-\rho }\rho ^{2n+k} n^{2}(\rho+
2(n+1)) }{n^{2}[(k+1)^{2}-4+(2n+k+1)(18n+3k+7)-4(n+1)^{2}(n+k+1)^{2}]}$}.
\end{eqnarray}%
On the other hand, according to the weak magnetic field condition, i.e, $%
\zeta ^{4}\ll 1$, the positive and negative energy eigenvalues are
\begin{equation}
\tilde{E}_{+}\approx M(1+2(n+1)\zeta ^{2})+O(\zeta ^{4}),  \label{42b}
\end{equation}%
\begin{equation}
\tilde{E}_{-}\approx -M(1+2n\zeta ^{2})+O(\zeta ^{4}),  \label{42bb}
\end{equation}%
and, then, the zeroth components of the current for the $\pm $ energy
eigenvalues become
\begin{equation}
\tilde{J}_{+}^{0}=\frac{{2\zeta ^{2}M}e^{-\rho }\rho ^{2n+k-1}(n+\rho +1)^{2}%
}{\pi \Gamma (2n+k)[(k+1)^{2}-4+(2n+k+1)(18n+3k+7)]},
\end{equation}
\begin{equation}
\tilde{J}_{-}^{0}=\frac{{2\zeta ^{2}M}e^{-\rho }\rho ^{2n+k-1}n^{2}}{\pi
\Gamma (2n+k)4(n+k+1)^{2}},
\end{equation}
respectively. Letting the currents, $J_{\pm }^{0}$ and $\tilde{J_{\pm }^{0}}$%
, the Noether charges of these currents are computed by the following
relations;
\begin{equation}
Q_{\pm }=\int (J^{0})_{\pm }dS_{0},
\end{equation}%
where the $Q_{\pm }$ are Noether charges and the $dS_{0}$ is a hypersurface
\cite{o1,gs1,zk}. If the currents, $J_{\pm }^{0}$ and $\tilde{J_{\pm }^{0}}$,
are integrated on the hypersurface, $dS_{0}=\rho d\rho d\theta $, we get the
Noether charges, respectively, as follows;
\begin{eqnarray}
\frac{Q_{+}}{M} &=\frac{2\zeta ^{2}}{\pi }  \nonumber \\
&\resizebox{.62\hsize}{!}{$\times\frac{ \left( 2n+1\right)+ (2n+k)(4n+k+3)
}{[(k+1)^{2}-4+(2n+k+1)(18n+3k+7)-4(n+k+1)^{2}]}$},\quad   \label{gg1}
\end{eqnarray}%

\begin{eqnarray}
\frac{Q_{-}}{M} &=\frac{2\zeta ^{2}}{\pi }  \nonumber \\
&\resizebox{.68\hsize}{!}{$\times\frac{n^{2}(2n+k)(4n+k+3)
}{n^{2}[(k+1)^{2}-4+(2n+k+1)(18n+3k+7)]-4(n+1)^{2}(n+k+1)^{2}}$},\quad
\label{gg2}
\end{eqnarray}%

\begin{eqnarray}
\frac{\tilde{Q_{+}}}{M}=\frac{2\zeta ^{2}((n+1)^{2}+(2n+k)(4n+k+3))}{\pi
((k+1)^{2}-4+(2n+k+1)(18n+3k+7))},
\end{eqnarray}%

\begin{eqnarray}
\frac{\tilde{Q_{-}}}{M}=\frac{\zeta ^{2}n^{2}}{2\pi (n+k+1)^{2}}.
\end{eqnarray}%
And, we plot these charges as follows;\qquad\qquad\qquad\qquad\qquad\qquad\qquad\qquad\qquad\qquad
\begin{figure}[h]
\begin{center}
\includegraphics[width=9cm, angle=0]{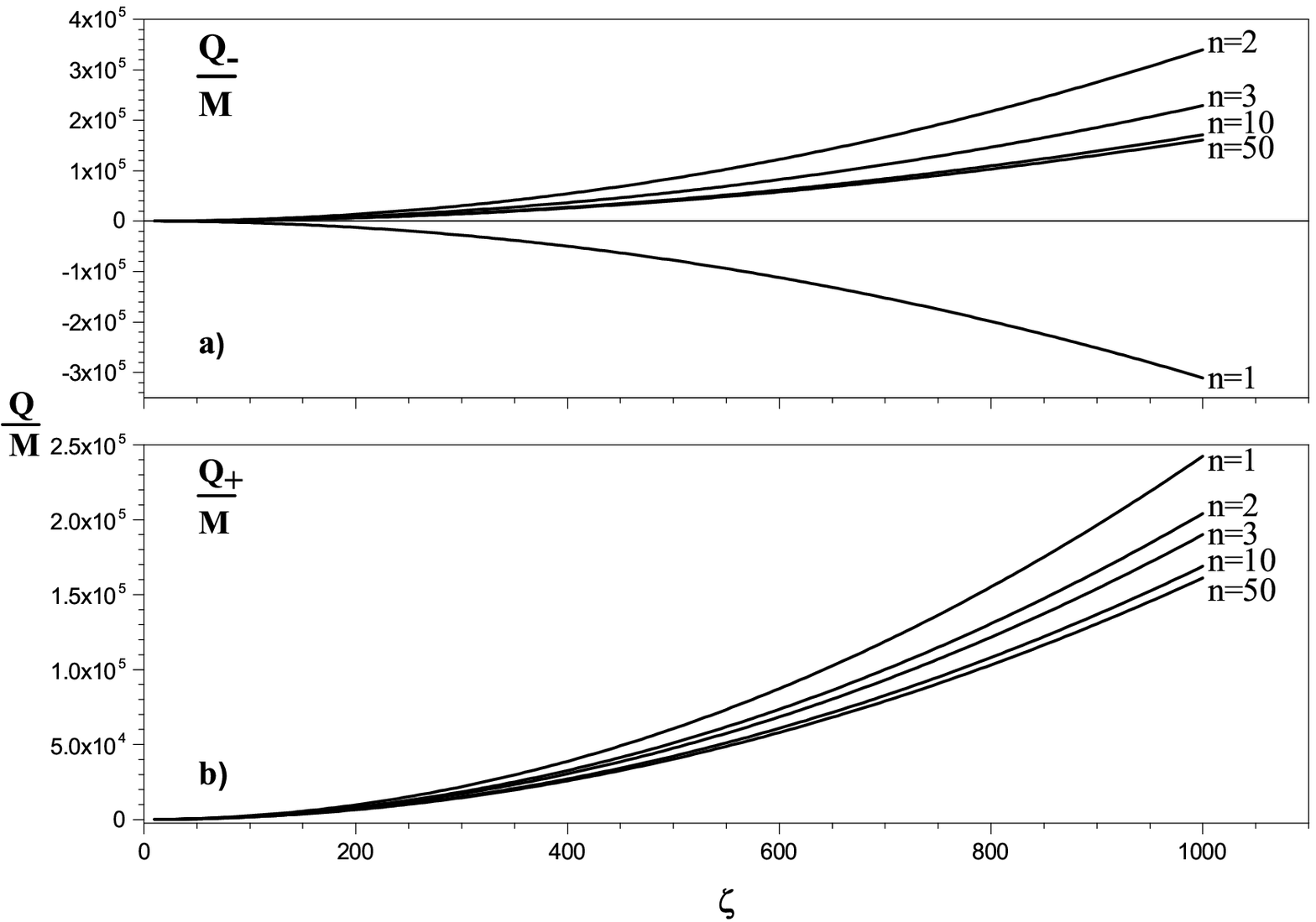}
\end{center}
\caption{The graph presents $\frac{Q_{\pm }}{M}$ as a function of $\protect\zeta $ ($%
\protect\zeta ^{4}\gg 1$, $k=\frac{3}{2}$).\newline
$\mathbf{a})$ The curve for $\frac{Q_{-}}{M}$ presents a negative increase for $n=1$, and it has a positive maximum  value for $n=2$. It also decreases
with increasing  values of $n$. \newline
$\mathbf{b})$ The $\frac{Q_{+}}{M}$ curve reaches to a maximum value for $n=1$.
 Increasing $n$ shows a similar behaviour as $\frac{Q_{-}}{M}$.  }
\label{fig1}
\end{figure}
\begin{figure}[!h]
\begin{center}
\includegraphics[width=9cm, angle=0]{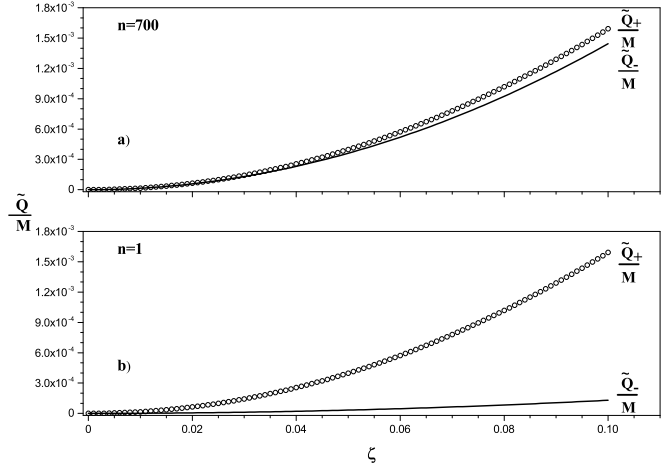}
\end{center}
\caption{The  $\frac{\tilde{Q_{\pm}}}{M}$  is plotted as a function of $\protect%
\zeta$ ( for $\protect\zeta^{4}\ll 1$, $k=\frac{3}{2}). The $ $\frac{\tilde{Q_{+}}%
}{M}$ curves for all of $n$ values get the same values. $\frac{\tilde{Q_{-}}}{M%
}$ curves  change with increasing $n$ values. Then
curves approach  to $\frac{\tilde{Q_{+}}}{M}$ curves. The curves merge for $n \geq 750$.\\
The  $\frac{\tilde{Q_{+}}}{M}$  is plotted as a function of $\protect
\zeta$ ( for $\protect\zeta^{4}\ll 1$, $k=\frac{3}{2})$ for $n=1$ and $n=700$ in respective panels.}
\label{fig2}
\end{figure}
We consider the behaviour of the proportion $\frac{Q}{M}$ according to the
strong and weak magnetic fields, see Figs: 1 and 2. From these figures, we see that the $\frac{Q_{+}}{M}$ values coincide with the $%
\frac{Q_{-}}{M}$ values as from $n\geq 10$ under the strong magnetic field
conditions while the $\frac{\tilde{Q_{+}}}{M}$ values coincide with the $%
\frac{\tilde{Q_{-}}}{M}$ values as from $n\geq 750$ under the weak magnetic
field conditions, but, under both conditions, the difference between the $%
\frac{Q_{+}}{M}$ values and the $\frac{Q_{-}}{M}$ values increase as the n
values get smaller. Also, given the Noether charge derived from the
probabilistic particle current, we can say that the magnetic field on the
particle production is more effective in the small $n$ values accoording to
the antiparticle production, but for the large $n$ values it is same.

\section{The spin-1 particle in the (2+1) dimensional curved spacetime}

\label{fieldeqns} \qquad To discuss the physical and mathematical features
of the spin-1 particle in a 2+1 dimensional spacetime curved background, the
relativistic quantum mechanical wave equation of the spin-1 particle is
easily generalized to a curved spacetime as follows;
\begin{equation}
\left[ \left( \overline{\mathbf{\sigma }}^{\mu }\left( x\right) \times
1+1\otimes \overline{\mathbf{\sigma }}^{\mu }\left( x\right) \right) \left(
P_{\mu }-i\Omega _{\mu }\right) -\left( 1\otimes 1\right) 2M\right] \Psi
\left( x\right) =0,  \label{51}
\end{equation}%
where$\ \overline{\mathbf{\sigma }}^{\mu }\left( x\right) $ are the
spacetime dependent Dirac matrices, $\Omega _{\mu }$ is the spin connection
of the spin-1 particle in 2+1 dimensions and its expression in terms of the
spin connection of the spin $-\frac{1}{2}$ particles, $\Gamma _{\mu }\left(
x\right) $, is written as
\begin{equation}
\Omega _{\mu }=\Gamma _{\mu }\left( x\right) \otimes 1+1\otimes \Gamma _{\mu
}\left( x\right).  \label{con}
\end{equation}
To derive solutions of the spin-1 particle wave equation in a curved
spacetime, as an example, we consider the following 2+1 dimensional
gravitational background \cite{a25,l1}:
\begin{equation}
ds^{2}=l^{2}\left[ d\tau ^{2}-\cosh ^{2}\tau \left( d\theta ^{2}+\sin
^{2}\theta \ d\phi ^{2}\right) \right] ,  \label{52}
\end{equation}%
where $\ \tau =t/l$ and $\tau \in \left( -\infty ,\infty \right) $, $\theta
\in \lbrack 0,\pi )$, $\phi \in \lbrack 0,2\pi )$ and $l$ is the radius of
the universe and related to the cosmological constant. $\Lambda$, as $l=%
\frac{1}{|\Lambda|}$. Also, the universe is a two spheres. It contracts to its
minimum area of $4\pi l^{2}$ at $\tau =0$ and expands again \cite{l1}. Then, the
metric tensor of the gravitational background and its inverse are written in
the following way;
\begin{eqnarray}
g_{\mu \nu } &=&{diag}\left( l^{2},-l^{2}\cosh ^{2}\tau ,-l^{2}\cosh
^{2}\tau \sin ^{2}\theta \right) ,  \nonumber \\
g^{\mu \nu } &=&{diag}\left( 1/l^{2},-1/l^{2}\cosh ^{2}\tau ,-1/l^{2}\cosh
^{2}\tau \sin ^{2}\theta \right) ,  \label{53}
\end{eqnarray}%
respectively, and the $g^{\mu \upsilon }$ is defined in terms of the triad
fields, $e_{i}^{\mu }\left( x\right) ,$\ as follows;

\begin{equation}
g^{\mu \upsilon }=e_{i}^{\mu }\left( x\right) e_{j}^{\nu }\left( x\right)
\eta ^{ij}.  \label{54}
\end{equation}
The triads of the curved background are explicitly written as

\begin{equation}
e_{i}^{\mu }\left( x\right) ={diag}\left( 1/l,1/l\cosh \tau ,1/l\cosh \tau
\sin \theta \right) .  \label{56}
\end{equation}%
The spacetime dependent Dirac matrices,\ $\overline{\mathbf{\sigma }}^{\mu
}\left( x\right) ,\ $are presented in terms of the triads and the constant
Dirac matrices, $\overline{\mathbf{\sigma }}^{i}$, as follows;

\begin{equation}
\overline{\mathbf{\sigma }}^{\mu }\ \left( x\right) =e_{i}^{\mu }\left(
x\right) \overline{\mathbf{\sigma }}^{i}.  \label{57}
\end{equation}%
And, the explicit form of the $\Gamma _{\mu }\left( x\right) $ is
\begin{equation}
\Gamma _{\mu }\left( x\right) =-\frac{1}{8}g_{\alpha \nu }\Gamma _{\beta \mu
}^{\nu }\left[ \overline{\mathbf{\sigma }}^{\alpha }\left( x\right) ,
\overline{\mathbf{\sigma }}^{\beta }\left( x\right) \right] ,  \label{58}
\end{equation}
where $\Gamma _{\beta \ \mu }^{\nu }$ are the Christoffel symbols, and also
its components in the curved background \cite{a25} are written as
\begin{eqnarray}
&\resizebox{.72\hsize}{!}{$\Gamma _{0}=0,\ \Gamma
_{1}=-\frac{1}{2}\overline{\mathbf{\sigma }}^{0}\overline{\mathbf{\sigma
}}^{1}\sinh \tau ,{\ }\Gamma _{2}=-\frac{1}{2}\left(
\overline{\mathbf{\sigma }}^{0}\overline{\mathbf{\sigma }}^{2}\sinh \tau
\sin \theta +\overline{\mathbf{\sigma }}^{1}\overline{\mathbf{\sigma
}}^{2}\cos \theta \right)$} .  \label{60}
\end{eqnarray}%
Then, Eq (\ref{51}) in the curved background is as follows;
\begin{equation}
\begin{array}{c}
\left( \partial _{\tau }+\tanh \tau +iMl\right) \psi _{+}+\frac{i}{\cosh
\tau }\left( \partial _{\theta }-i\frac{\partial _{\phi }}{\sin \theta }%
\right) \psi _{0}=0, \\
\resizebox{.72\hsize}{!}{$\frac{i}{\cosh \tau }\left( \partial _{\theta
}+i\frac{\partial _{\phi }}{\sin \theta }+\cot \theta \right) \psi
_{+}+\frac{i}{\cosh \tau }\left( \partial _{\theta }-i\frac{\partial _{\phi
}}{\sin \theta }+\cot \theta \right) \psi _{-}=-2iMl\psi _{0}$}, \\
\frac{i}{\cosh \tau }\left( \partial _{\theta }+i\frac{\partial _{\phi }}{%
\sin \theta }\right) \psi _{0}-\left( \partial _{\tau }+\tanh \tau
-iMl\right) \psi _{-}=0.%
\end{array}
\label{61}
\end{equation}

To find the general solution, we use the separation of variable method. In
this connection, the rising and lowering operators of the spin-1 particle, $%
\partial _{\pm }$, are defined as
\begin{equation}
\partial _{\pm }=\left( \mp \partial _{\theta }+i\frac{\partial _{\phi }}{%
\sin \theta }+\frac{1}{2}\left( \overline{\mathbf{\sigma }}^{0}\otimes
1+1\otimes \overline{\mathbf{\sigma }}^{0}\right) \cot \theta \right) .
\label{62}
\end{equation}%
and their eigenfunctions in terms of the rotation group $d_{\lambda
,m}^{j}\left( \theta \right) $ are expressed as
\begin{equation}
D_{\lambda ,m}^{j}\left( \theta ,\phi \right) =\langle \lambda |R\left(
\theta ,\phi \right) |jm\rangle =e^{i\lambda \phi }d_{\lambda ,m}^{j}\left(
\theta \right).  \label{63}
\end{equation}%
And, the irreducible representations of the rotation group $d_{\lambda
,m}^{j}\left( \theta \right) $ are given by%
\begin{eqnarray*}
d_{\lambda ,m}^{j}\left( \theta \right) &=\frac{\left( -1\right) ^{j-m}}{%
\left( \lambda +m\right) !}\sqrt{\frac{\left( j+\lambda \right) !\left(
j+m\right) !}{\left( j-\lambda \right) !\left( j-m\right) !}}\sin \left(
\theta /2\right) ^{2j}\cot \left( \theta /2\right) ^{\lambda +m} \\
&\times _{2}F_{1}\left( \lambda -j,m-j;\ \lambda +m+1;\ -\cot ^{2}\left(
\theta /2\right) \right)
\end{eqnarray*}%
\cite{p11}. When the separation of variable method is employed,  $\Psi \left(
\tau,\theta ,\phi \right) $ can be expanded in terms of the angular momentum eigenfunctions
as%
\begin{equation}
\Psi \left( \tau ,\theta ,\phi \right) =4\pi {\sum_{jm} }\frac{\left(
2j+1\right) }{\cosh \tau }\left(
\begin{array}{c}
F_{+}\left( \tau \right) D_{+1,m}^{j}\left( \theta ,\phi \right) \\
F_{0}\left( \tau \right) D_{0,m}^{j}\left( \theta ,\phi \right) \\
F_{-}\left( \tau \right) D_{-1,m}^{j}\left( \theta ,\phi \right)%
\end{array}%
\right)  \label{64}
\end{equation}%
and the $\partial _{\pm }$ operators act on $D_{\lambda ,m}^{j}\left( \theta
,\phi \right) $ as follows;%
\begin{equation}
\partial _{\pm }D_{\lambda ,m}^{j}\left( \theta ,\phi \right) =\sqrt{\left(
j\pm \lambda +1\right) \left( j\mp \lambda \right) }D_{\lambda \pm
1,m}^{j}\left( \theta ,\phi \right) .  \label{65}
\end{equation}%
Then, Eq (\ref{61}) is rewritten as%
\begin{equation}
\begin{array}{c}
\left( iMl+\partial _{\tau }\right) F_{+}\left( \tau \right) -\frac{i}{\cosh
\tau }\sqrt{j\left( j+1\right) }F_{0}\left( \tau \right) =0, \\
\left( iMl-\partial _{\tau }\right) F_{-}\left( \tau \right) +\frac{i}{\cosh
\tau }\sqrt{j\left( j+1\right) }F_{0}\left( \tau \right) =0, \\
MlF_{0}\left( \tau \right) -\frac{1}{2\cosh \tau }\sqrt{j\left( j+1\right) }%
\left( F_{-}\left( \tau \right) -F_{+}\left( \tau \right) \right) =0.%
\end{array}
\label{67}
\end{equation}%
From these equations, the $F_{0}\left( \tau \right) $ can be defined as%
\begin{equation}
F_{0}\left( \tau \right) =\frac{1}{2Ml\cosh \tau }\sqrt{j\left( j+1\right) }%
\left( F_{-}\left( \tau \right) -F_{+}\left( \tau \right) \right) .
\label{68}
\end{equation}%
and substituting it into the first and second rows of Eq (\ref{67}), we get
the following equations for the $F_{\pm }\left( \tau \right)$ :%
\begin{equation}
\left[ \partial _{\tau }^{2}+2\left( \tanh \tau \right) \partial _{\tau }\pm
2iMl\tanh \tau +M^{2}l^{2}+\frac{j\left( j+1\right) }{\cosh ^{2}\tau }\right]
F_{\pm }\left( \tau \right) =0.  \label{69}
\end{equation}%
These differential equations are satisfied the following solutions \cite{I5}%
:
\begin{equation}
\resizebox{.68\hsize}{!}{$F_{+}\left( \tau \right) =\left( \frac{1-\tanh
\tau }{2}\right) ^{j+1}\left[ C_{+}\xi \left( \tau \right)
^{-\frac{iMl}{2}}F_{1}\left( \tau \right) +D_{-}\xi \left( \tau \right)
^{\frac{iMl}{2}+1}F_{2}\left( \tau \right) \right]$} ,  \label{70}
\end{equation}

\begin{equation}
\resizebox{.68\hsize}{!}{$F_{-}\left( \tau \right) =-\left( \frac{1-\tanh
\tau }{2}\right) ^{j+1}\left[ D_{+}\xi \left( \tau \right)
^{-\frac{iMl}{2}+1}F_{3}\left( \tau \right) +C_{-}\xi \left( \tau \right)
^{\frac{iMl}{2}}F_{4}\left( \tau \right) \right]$},  \label{71}
\end{equation}%
where the $F_{i}\left( \tau \right) $ functions are an abbreviations of the
Hypergeometric functions as follows;
\[
\begin{array}{c}
F_{1}\left( \tau \right) =\ _{2}F_{1}\left( -j-iMl,\ -j-1;\ -iMl;\ -\xi
\left( \tau \right) \right) , \\
F_{2}\left( \tau \right) =\ _{2}F_{1}\left( -j+iMl,\ -j+1\ ;\ 2+iMl;\ -\xi
\left( \tau \right) \right) , \\
F_{3}\left( \tau \right) =\ _{2}F_{1}\left( -j-iMl,\ -j+1\ ;\ 2-iMl;\ -\xi
\left( \tau \right) \right) , \\
F_{4}\left( \tau \right) =\ \ _{2}F_{1}\left( -j+iMl,\ -j-1;\ iMl;\ -\xi
\left( \tau \right) \right)
\end{array}%
\]%
and $\xi \left( \tau \right) =\frac{1+\tanh \tau }{1-\tanh \tau }.$ Also, the
coefficients $C_{\pm }$ and $D_{\pm }$ are related each other by the
following relations:%
\begin{eqnarray}
C_{\pm }=-\frac{\left( Ml\pm \frac{i}{2}\right) ^{2}+\frac{1}{4}}{\left( j+%
\frac{1}{2}\right) ^{2}-\frac{1}{4}}D_{\pm }
\end{eqnarray}
To construct the wave function when $\tau $ goes $-\infty $, we consider the
asymptotic forms of $F_{+}\left( \tau \right) ,$ $F_{0}\left( \tau \right) $
and $F_{-}\left( \tau \right) $ in Eq (\ref{68}), Eq (\ref{70}) and Eq (\ref%
{71}), then, the wave function is written as

\begin{eqnarray}
\resizebox{.10\hsize}{!}{$\Psi \left( \tau ,\theta ,\phi \right) $}%
=\sum_{jm}2\sqrt{2j+1}e^{\tau }\left(
\begin{array}{c}
\resizebox{.35\hsize}{!}{$\left(C_{+}e^{-iMl\tau }+e^{2\tau }D_{-}e^{iMl\tau
}\right) D_{+1,m}^{j}\left( \theta ,\phi \right)$} \\
-\frac{\sqrt{j\left( j+1\right) }}{Ml}e^{\tau }\{ \verb|F|_{+}e^{-iMl\tau}
\\
+\verb|F|_{-}e^{iMl\tau }\} D_{0,m}^{j}\left( \theta ,\phi \right) \\
-\resizebox{.35\hsize}{!}{$\left(e^{2\tau } D_{+}e^{-iMl\tau }+
C_{-}e^{iMl\tau }\right) D_{-1,m}^{j}\left( \theta ,\phi \right)$}%
\end{array}%
\right),\qquad  \label{78}
\end{eqnarray}
where $\verb|F|_{+}= C_{+}+e^{2\tau } D_{+}$ and $\verb|F|_{-}=
C_{-}+e^{2\tau }D_{-}.$ The asymptotic expressions of the particle $\Psi
^{+} $ and antiparticle $\Psi ^{-}\ $solutions can be written separately as
the following
\begin{eqnarray}
\Psi ^{+}=2\sqrt{2j+1}e^{\tau }\left(
\begin{array}{c}
C_{+}D_{+1,m}^{j}\left( \theta ,\phi \right) \\
-\frac{\sqrt{j\left( j+1\right) }}{Ml}e^{\tau }(C_{+}+e^{2\tau}D_{+}
)D_{0,m}^{j}\left( \theta ,\phi \right) \\
-D_{+}e^{2\tau }D_{-1,m}^{j}\left( \theta ,\phi \right).%
\end{array}%
\right) e^{-iMl\tau },  \label{74}
\end{eqnarray}

\begin{eqnarray}
\Psi ^{-}=2\sqrt{2j+1}e^{\tau }\left(
\begin{array}{c}
D_{-}\ e^{2\tau }D_{+1,m}^{j}\left( \theta ,\phi \right) \\
-\frac{\sqrt{j\left( j+1\right) }}{Ml}e^{\tau
}(C_{-}+e^{2\tau}D_{-})D_{0,m}^{j}\left( \theta ,\phi \right) \\
-C_{-}D_{-1,m}^{j}\left( \theta ,\phi \right)%
\end{array}%
\right) e^{iMl\tau },  \label{75}
\end{eqnarray}
where $C_{+}$ and $C_{-}$ are normalization coefficients, and they are
calculated as
\begin{equation}
\resizebox{.76\hsize}{!}{$\left\vert C_{+}\right\vert =\left\vert
C_{-}\right\vert =\frac{1}{l} \left[ 1-\frac{j\left( j+1\right)
}{M^{2}l^{2}}e^{2\tau }+3\frac{j^{2}\left( j+1\right) ^{2}}{M^{2}l^{2}\left(
1+M^{2}l^{2}\right) }e^{4\tau }-\frac{j^{3}\left( j+1\right)
^{3}}{M^{4}l^{4}\left( 1+M^{2}l^{2}\right) }e^{6\tau }\right] ^{-1/2}$}. \label{75ab}
\end{equation}

To discuss the particle creation in this background, we construct the
particle and the antiparticle solutions as $\tau \rightarrow \infty .$
As $\tau $ goes to $\infty $, the wave function is

\begin{equation}
\resizebox{.11\hsize}{!}{$\tilde\Psi \left( \tau ,\theta ,\phi \right)$}%
=\sum_{jm}2\sqrt{2j+1}e^{-\tau }\left(
\begin{array}{c}
\resizebox{.33\hsize}{!}{$\left( \tilde C_{+}e^{-iMl\tau }+e^{-2\tau }\tilde
D_{-}e^{iMl\tau }\right) D_{+1,m}^{j}\left( \theta ,\phi \right)$} \\
-\frac{1}{Ml}\sqrt{j\left( j+1\right) }e^{-\tau }\{\tilde{\verb|F|_{+}}%
e^{-iMl\tau } \\
+\tilde{\verb|F|_{-}}e^{iMl\tau }\}D_{0,m}^{j}\left( \theta ,\phi \right)
\\
-\resizebox{.33\hsize}{!}{$\left(e^{-2\tau } \tilde D_{+}e^{-iMl\tau
}+\tilde C_{-}e^{iMl\tau }\right) D_{-1,m}^{j}\left( \theta ,\phi \right)$}%
\end{array}%
\right) ,  \label{82}
\end{equation}%
where $\tilde{\verb|F|_{+}}=\tilde{C}_{+}+e^{-2\tau }\tilde{D}_{+}$ and $%
\tilde{\verb|F|_{-}}=\tilde{C}_{-}+e^{-2\tau }\tilde{D}_{-}.$ The asymptotic
expressions of the particle $_{+}\Psi $ and anti-particle $_{-}\Psi \ $%
solution to be written separately as the following
\begin{equation}
_{+}\tilde{\Psi}=2\sqrt{2j+1}e^{-\tau }\left(
\begin{array}{c}
\tilde{C}_{+}D_{+1,m}^{j}\left( \theta ,\phi \right)  \\
\resizebox{.38\hsize}{!}{$-\frac{\sqrt{j\left( j+1\right) }}{Ml}e^{-\tau
}(\tilde C_{+}+e^{-2\tau }\tilde D_{+})D_{0,m}^{j}\left( \theta ,\phi
\right)$} \\
-e^{-2\tau }\tilde{D}_{+}D_{-1,m}^{j}\left( \theta ,\phi \right).
\\
\end{array}%
\right) e^{-iMl\tau },\qquad   \label{76}
\end{equation}%
\begin{equation}
_{-}\tilde{\Psi}=2\sqrt{2j+1}e^{-\tau }\left(
\begin{array}{c}
\tilde{D}_{-}\ e^{-2\tau }D_{+1,m}^{j}\left( \theta ,\phi \right)  \\
\resizebox{.37\hsize}{!}{$-\frac{\sqrt{j\left( j+1\right) }}{Ml}e^{-\tau
}(\tilde C_{-}+e^{-2\tau }\tilde D_{-})D_{0,m}^{j}\left( \theta ,\phi
\right)$} \\
-\tilde{C}_{-}D_{-1,m}^{j}\left( \theta ,\phi \right)
\end{array}%
\right) e^{iMl\tau },\qquad  \label{76z}
\end{equation}%
where $\tilde{C}_{+}=a_{1}C_{+}$, $\tilde{C}_{-}=b_{2}C_{-}$, $\tilde{D}%
_{+}=b_{1}D_{+}$ and $\tilde{D}_{-}=a_{2}D_{-}.$ Here $\tilde{C}_{+}$ and $%
\tilde{C}_{-}$ are normalization coefficients and the $a_{1}$, $b_{1}$, $a_{2}$
and $b_{2}$ are
\[
\begin{array}{c}
a_{1}=\frac{\Gamma \left( -iMl\right) \Gamma \left( -iMl+1\right) }{\Gamma
\left( -iMl-j\right) \Gamma \left( -iMl+j+1\right) },\quad a_{2}=\frac{%
\Gamma \left( 2+iMl\right) \Gamma \left( iMl-1\right) }{\Gamma \left(
-j+iMl\right) \Gamma \left( j+iMl+1\right) }, \\
\\
b_{1}=\frac{\Gamma \left( 2-iMl\right) \Gamma \left( -iMl-1\right) }{\Gamma
\left( -iMl-j\right) \Gamma \left( -iMl+j+1\right) },\quad b_{2}=\frac{%
\Gamma \left( iMl\right) \Gamma \left( iMl+1\right) }{\Gamma \left(
-j+iMl\right) \Gamma \left( j+iMl+1\right) }.%
\end{array}%
\]%
Letting Eq (\ref{76}) and Eq (\ref{76z}), we obtain the normalization
coefficients as follows
\begin{equation}
\resizebox{.79\hsize}{!}{$\left\vert\tilde C_{+}\right\vert
=\left\vert\tilde C_{-}\right\vert =\frac{1}{l} \left[ 1-\frac{j\left(
j+1\right) }{M^{2}l^{2}}e^{-2\tau }+3\frac{j^{2}\left( j+1\right)
^{2}}{M^{2}l^{2}\left( 1+M^{2}l^{2}\right) }e^{-4\tau }-\frac{j^{3}\left(
j+1\right) ^{3}}{M^{4}l^{4}\left( 1+M^{2}l^{2}\right) }e^{-6\tau }\right]
^{-\frac{1}{2}}$}.
\end{equation}

The current for the spin-1 particle in 2+1 dimensional curved spacetime can
be computed by using Eq (\ref{20}) and Eq (\ref{23}). Therefore, using
Eq (\ref{74}), Eq (\ref{75}), Eq (\ref{75ab}) and Eq (\ref{23}), the zeroth component of the particle current, $(J^{0}_{+ })$, and the antiparticle current, $(J^{0}_{-})$, are obtained by the following away;

\begin{eqnarray}
(J^{0})_{\pm } &=\pm \frac{1}{l}4e^{2\tau }\left( 2j+1\right) \left\{
|C_{\pm }\right\vert ^{2}D_{\pm 1,m}^{j}\left( \theta ,\phi \right) ^{\ast
}D_{\pm 1,m}^{j}\left( \theta ,\phi \right)   \nonumber \\
&-\left\vert D_{\pm }\right\vert ^{2}e^{4\tau }D_{\mp 1,m}^{j}\left( \theta
,\phi \right) ^{\ast }D_{\mp 1,m}^{j}\left( \theta ,\phi \right)
\},\allowbreak
\end{eqnarray}%
and, using the orthogonality relation for the rotation group \cite{p11}, then, the
Noether charges from these currents are calculated  as

\begin{eqnarray}
Q_{\pm } &=\int (J^{0})_{\pm }dS_{0}  \nonumber \\
&=\pm \frac{1}{l}\left[ 1-\frac{j^{2}\left( j+1\right) ^{2}}{%
M^{2}l^{2}\left( 1+M^{2}l^{2}\right) }e^{4\tau }\right]   \nonumber \\
&\times \resizebox{.70\hsize}{!}{$ \left[ 1-\frac{j\left( j+1\right)
}{M^{2}l^{2}}e^{2\tau }+3\frac{j^{2}\left( j+1\right) ^{2}}{M^{2}l^{2}\left(
1+M^{2}l^{2}\right) }e^{4\tau }-\frac{j^{3}\left( j+1\right)
^{3}}{M^{4}l^{4}\left( 1+M^{2}l^{2}\right) }e^{6\tau }\right] ^{-1}$}
\nonumber \\
&\cong \pm \frac{1}{l}\left( 1+\frac{j\left( j+1\right) }{4M^{2}l^{2}}%
e^{2\tau }+O\left( e^{4\tau }\right) \right) ,  \label{79}
\end{eqnarray}%
where $dS_{0}=l^{2}cosh^{2}\tau \sin \theta d\theta d\phi $. In the same
way, from Eq (\ref{76}), Eq (\ref{76z}) and Eq (\ref{23}), the zeroth components of currents of particle $(\tilde{J}^{0}_{+})$
and antiparticle $(\tilde{J}^{0}_{-})$  as $\tau$ goes to $\infty$ are obtained as the following way:

\begin{eqnarray}
(\tilde J^{0})_{\pm}&=\pm\frac{1}{l}4e^{-2\tau }\left( 2j+1\right)
\left\{\vert \tilde C_{\pm}\right\vert ^{2}D_{\pm1,m}^{j}\left( \theta ,\phi
\right) ^{\ast }D_{\pm1,m}^{j}\left( \theta ,\phi \right)  \nonumber \\
&- \left\vert\tilde D_{\pm}\right\vert ^{2}e^{-4\tau }
D_{\mp1,m}^{j}\left( \theta ,\phi \right) ^{\ast } D_{\mp1,m}^{j}\left(
\theta ,\phi \right) \}, \allowbreak
\end{eqnarray}
and, from these currents, the Noether charges for the particle and
antiparticle are calculated as

\begin{eqnarray}
\tilde{Q}_{\pm } &=\int (\tilde{J}^{0})_{\pm }dS_{0}  \nonumber \\
&=\pm \frac{1}{l}\left[ 1-\frac{j^{2}\left( j+1\right) ^{2}}{%
M^{2}l^{2}\left( 1+M^{2}l^{2}\right) }e^{-4\tau }\right]  \nonumber \\
&\times \resizebox{.70\hsize}{!}{$ \left[ 1-\frac{j\left( j+1\right)
}{M^{2}l^{2}}e^{-2\tau }+3\frac{j^{2}\left( j+1\right)
^{2}}{M^{2}l^{2}\left( 1+M^{2}l^{2}\right) }e^{-4\tau }-\frac{j^{3}\left(
j+1\right) ^{3}}{M^{4}l^{4}\left( 1+M^{2}l^{2}\right) }e^{-6\tau }\right]
^{-1}$}  \nonumber \\
&\cong \pm \frac{1}{l}\left( 1+\frac{j\left( j+1\right) }{4M^{2}l^{2}}%
e^{-2\tau }+O\left( e^{-4\tau }\right) \right) .  \label{79}
\end{eqnarray}%
These results show that as $\tau $ goes to $\pm \infty $, $\frac{\tilde{Q}_{\pm }%
}{|\Lambda |}$ and $\frac{Q_{\pm }}{|\Lambda |}$, respectively, converge to $%
\pm 1$. This means that the universe only is composed of the $\pm $ $%
|\Lambda |$ vacuum energies in its beginning and ending time.

\section{Concluding remarks}

\label{conc}

\qquad In this study, we  have introduced a relativistic quantum mechanical wave
equation of the spin-1 particle as an excited state of the zitterbewegung.
This wave function has three independent components. And letting a complex
vector potential, $A^{\mu }$, and fields, $F^{\mu \nu },$ in terms of three
components, we show that the equation is consistent with the 2+1 dimensional
Proca theory. At the same time, we see that this equation has two
eigenstates, particle and antiparticle states or negative and positive
energy eigenstates, respectively, in the rest frame and satisfy $SO(2,1)$
spin algebra. Apart from the free particle solution of the equation, we have
derived the exact solutions of it in presence of the constant magnetic field
and the curved background. From these solutions, we have constructed the current
components and observed a spin-1 particle current in presence  of a
constant magnetic field that decreases by $e^{-\rho },$ while the spin-1
particle current in the curved spacetime oscillates in time, which means
that there are temporary particle creations. On the other hand, from the
currents, we evaluate the Noether charges. And, we see that the $\frac{Q_{+}}{%
M}$ values coincide with the $\frac{Q_{-}}{M}$ values as from $n\geq 10$
under the strong magnetic field conditions, while the $\frac{\tilde{Q_{+}}}{M%
}$ values coincide with the $\frac{\tilde{Q_{-}}}{M}$ values as from $n\geq
750$ under the weak magnetic field conditions, in this situation, the presence of
magnetic field triggers the  particle production more than antiparticle production in the small $n$ values,
but in the large values, the particle and antiparticle production are same. Also, in the curved
background, as $\tau $ goes to $\pm \infty $, the $\frac{Q_{\pm }}{|\Lambda |}
$ and $\frac{Q_{\pm }}{|\Lambda |}$, respectively, converge to $\pm 1$. So,
it can be said that, in the beginning and ending of the time, the universe may have,
completely, been composed of the particle and antiparticle with the positive and
negative $|\Lambda |$ vacuum energies, respectively.

Finally, besides to the results obtained in this study, we think that the consistent spin-1 particle wave
equation in the 2+1 dimensional spacetime gains much interest in physical area \cite{gy}.

\section*{Acknowledgments}

This work was supported by the Scientific Research Projects Unit of Akdeniz
University.


\end{document}